# Alfvénic Thermospheric Upwelling in a Global Geospace Model

Benjamin Hogan[1,2,†] 　William Lotko[1,2] 　and Kevin Pham[2]

[1]Thayer School of Engineering, Dartmouth College, Hanover, New Hampshire, USA, [2]High Altitude Observatory, National Center for Atmospheric Research, Boulder, CO, USA, [†]Now at Laboratory for Atmospheric and Space Physics, University of Colorado Boulder, Boulder, CO, USA and Department of Aerospace Sciences, University of Colorado Boulder, Boulder, CO, USA

**Contents**

**Correspondence**

B. Hogan
Benjamin.Hogan-1@colorado.edu

W. Lotko
wlotko@dartmouth.edu

K. Pham
phamkh@ucar.edu

**Abstract**　Motivated by low-altitude cusp observations of small-scale (~ 1 km) field-aligned currents (SSFACs) interpreted as ionospheric Alfvén resonator modes, we have investigated the effects of Alfvén wave energy deposition on thermospheric upwelling and the formation of air density enhancements in and near the cusp. Such density enhancements were commonly observed near 400 km altitude by the CHAMP satellite. They are not predicted by empirical thermosphere models, and they are well-correlated with the observed SSFACs. A parameterized model for the altitude dependence of the Alfvén wave electric field, constrained by CHAMP data, has been developed and embedded in the Joule heating module of the National Center for Atmospheric Research (NCAR) Coupled Magnetosphere-Ionosphere-Thermosphere (CMIT) model. The CMIT model was then used to simulate the geospace response to an interplanetary stream interaction region (SIR) that swept past Earth on 26-27 March 2003. CMIT diagnostics for the thermospheric mass density at 400 km altitude show: 1) CMIT without Alfvénic Joule heating usually underestimates CHAMP's *orbit-average* density; inclusion of Alfvénic heating modestly improves CMIT's orbit-average prediction of the density (by a few %), especially during the more active periods of the SIR event. 2) The improvement in CMIT's *instantaneous* density prediction with Alfvénic heating included is more significant (up to 15%) in the vicinity of the cusp heating region, a feature that the MSIS empirical thermosphere model misses for this event. Thermospheric density changes of 20-30 % caused by the cusp-region Alfvénic heating sporadically populate the polar region through the action of co-rotation and neutral winds.

## 1. Introduction

Thermospheric density anomalies with average values 20-30% larger than empirical model predictions are encountered by satellites orbiting near 400 km altitude in the high-latitude dayside thermosphere (Liu et al., 2005, Rentz & Lühr et al., 2008). About 40% of CHAMP satellite orbits exhibit such anomalies (Kervalishvili & Lühr, 2014). Soft electron precipitation, also referred to as broadband precipitation with energies of 100s eV (Živković et al, 2015), and intense, small-scale field-aligned currents (SSFACs) (Lühr, et al, 2004; Kervalishvili & Lühr, 2013) are well correlated with the anomalies in the cusp region. Enhanced Joule heating by quasistatic electric fields may contribute to the thermospheric upwelling that produces density anomalies, but observations (Schlegel et al., 2005) and models (Zhang et al., 2012; Deng et al., 2013) indicate that this mechanism alone is insufficient.

Simulation studies of the thermosphere (Brinkman et al., 2016), the coupled ionosphere-thermosphere (IT) (Deng et al, 2013) and the coupled magnetosphere-ionosphere-thermosphere (MIT) (Zhang et al., 2012, 2015a) show that soft electron precipitation in the cusp, combined with sufficiently intense, quasistatic electric fields, can produce density anomalies resembling those observed in the cusp region. In all of these studies, the soft electron precipitation augments the *F*-region Pedersen conductivity and the specific Joule heating rate at altitudes where the ambient air density is exponentially decreasing. The added heat and attendant increase in thermospheric temperature is effective in producing upwelling and a major density perturbation of the neutral gas (Jee et al., 2008; Clemmons et al., 2008; Brinkman et al., 2016).

Lotko and Zhang (2018) showed that the Joule heating rate resulting from persistent and more or less continuously driven, small-scale Alfvénic perturbations representative of the SSFACs regularly observed by CHAMP in the cusp can be significant at *F*-region altitudes (200-400 km). Transient Alfvénic perturbations accompanying discontinuities in solar wind driving are also expected to augment *F*-region Joule heating (Tu et al., 2011) and contribute to enhancements in thermospheric density at these altitudes, but isolated transients in Alfvénic energy deposition cannot account for the relatively high probability of observing a thermospheric density anomaly on any given satellite pass through the cusp. In contrast with quasistatic electric fields, which are practically independent of altitude in the thermosphere, Alfvénic electric fields are more intense in the *F*-region than in the *E*-region owing to the influence of the ionospheric Alfvén resonator (IAR). The IAR traps Alfvén waves and intensifies their amplitudes near the *F*-region peak in plasma density where the wave speed is slow (Trakhtengerts & Fel'Dshtein, 1984; Lysak, 1991). The results suggest that Alfvén wave energy deposition in the IAR may contribute to air density anomalies at CHAMP altitudes in the absence of Joule heating by intense quasistatic electric fields.

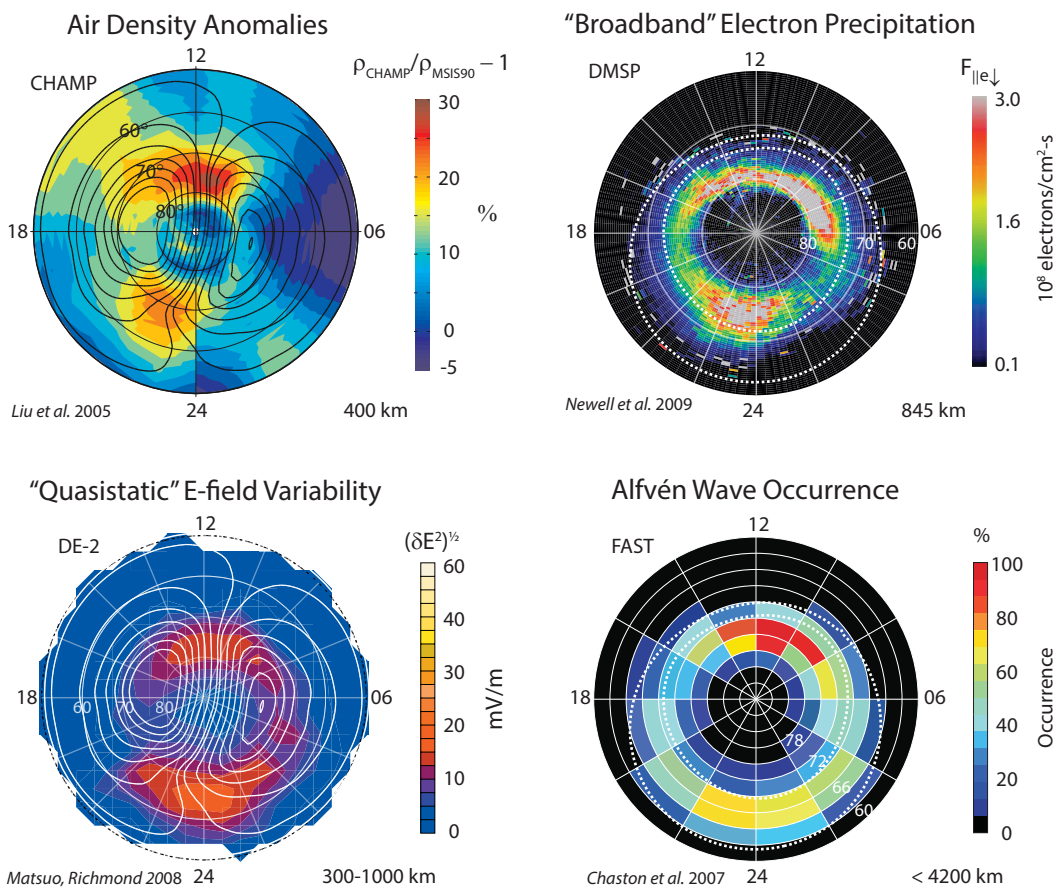

**Figure 1.** Top left: One-year average difference between air density recorded by CHAMP in the northern hemisphere and estimated from MSIS. Illustrative CMIT convection streamlines superposed. Top right: Statistical number flux of broadband electron precipitation from DMSP. Feldstein statistical auroral oval superposed as dotted lines. Bottom left: Statistical root-mean-square electric field variability from DE-2, with same illustrative convection lines. Bottom right: Occurrence of Alfvén waves recorded by FAST, also with Feldstein oval superposed. All figures are in MLT-MLAT (magnetic local time, magnetic latitude) coordinates. Source of plot and nominal altitude of measurements are indicated at the bottom of each.

The SSFACs recorded by CHAMP in association with density anomalies have been interpreted as IAR modes (Rother et al., 2007). However, distinguishing Alfvén waves with characteristic wave impedance and phase relations (e.g., Ishii et al., 1992) from quasistatic variability requires simultaneous electric and magnetic field measurements. CHAMP carried a sensitive magnetometer but no electric field sensor. Measurements from the Polar (Keiling et al., 2003) and FAST (Chaston et al., 2007; Hatch et al., 2017) satellites have been used to definitively identify Alfvénic fluctuations and their distributions in magnetic local time (MLT) and magnetic latitude (MLAT). The statistical distribution of the most probable occurrence of Alfvénic fluctuations from FAST shown in Figure 1 (lower right) correlate well with the MLT-MLAT distribution of CHAMP thermospheric density anomalies (Fig. 1, upper left), as does the distribution of soft electron precipitation derived from DMSP satellite measurements (Fig. 1, upper right).

The MLT-MLAT distribution of quasistatic electric field variability is also shown in Fig. 1 (lower left) because it has been posited as a possible causal agent for the enhanced Joule heating needed to produce the CHAMP density anomalies (Deng et al., 2013; Brinkman et al., 2016). This particular distribution derived by Matsuo & Richmond (2008) from DE-2 satellite data for summer conditions and clock angles of 45°-135° of the interplanetary magnetic field (IMF) also resembles the distribution of CHAMP anomalies. Distributions of electric field variability for winter and equinox conditions and other IMF clock angles are very different and, for some conditions (e.g., summer and IMF clock angles between 135°-225) exhibit very low amplitudes where CHAMP anomalies occur. Matsuo & Richmond's distributions of electric field variability assume the observed variation is spatial (no time variation) with contributions integrated over horizontal scales between 3 km and 500 km. The assumption of static electric field variability may limit the applicability of these results, particularly for contributions at the smaller scales (< few 100 km), for which Alfvénic rather than quasistatic variability is usually prominent (Ishii et al., 1992; Chaston et al., 2003; Lühr et al., 2015; Park et al., 2017).

We focus in this paper on the effects of Alfvén wave energy deposition on the thermospheric mass density in and near the low-altitude cusp at the nominal CHAMP altitude of 400 km. The Alfvén wave model of Lotko & Zhang (2018) is adapted as a parameterized model in the NCAR Coupled Magnetosphere-Ionosphere-Thermosphere (CMIT), which is used to simulate and analyze the MIT response to the stream interaction region (SIR) that swept past Earth on 26-27 March 2003. The relative importance of Alfvénic Joule heating for this event is characterized as a difference in thermospheric mass densities at 400 km altitude with and without inclusion of Alfvénic heating. The CMIT model includes i) a cusp-finding algorithm (Zhang et al., 2013), which dynamically determines where the Alfvénic Joule heating is to be applied, and ii) a causally regulated empirical specification for soft electron precipitation. The soft precipitation is modeled as both direct-entry cusp precipitation and Alfvén wave-induced "broadband" precipitation (Zhang et al., 2015b). As mentioned above, soft precipitation and some type of electric field enhancement are important factors in producing thermospheric density enhancement in the cusp region.

We first provide an overview of the CMIT model and the interplanetary conditions for the SIR event to be simulated (Section 2.1). The parameterized Alfvén wave model to be embedded in CMIT is described (Section 2.2), along with specification of the ambient IT parameters that determine the characteristics of Alfvén propagation and absorption (Section 2.3) and integration of the Alfvén wave heating model into CMIT (Section 2.4). Discussion of the simulation results (Section 3) and principal conclusions (Section 4) follow.

## 2. Methods

### 2.1 Overview of CMIT and the 26-27 March 2003 SIR event

NCAR's CMIT model has been used previously to simulate the MIT response to solar wind and interplanetary driving (Wang et al., 2004; Wiltberger et al., 2004; Zhang et al., 2012, 2015a, Liu et al., 2018). The magnetospheric component of CMIT is the Lyon-Fedder-Mobarry (LFM) global magnetosphere simulation model, which describes the coupling between the solar wind and magnetosphere by solving the three-dimensional equations of ideal magnetohydrodynamics (Lyon et al., 2004). The so-called "double-resolution" version (53x48x64 grid cells) of LFM is used here. Its physical domain is a distorted spherical grid extending 30 Earth radii ($R_E$) upstream from Earth into the solar wind, to 300 $R_E$ downstream on the nightside, and 100 $R_E$ on the sides. The LFM model is driven by time series of state variables for the solar wind and IMF obtained from interplanetary satellite measurements (the one-minute OMNI-combined data available at https://cdaweb.gsfc.nasa.gov/index.html/).

The ionosphere-thermosphere (IT) component of CMIT is modeled by the Thermosphere-Ionosphere Electrodynamics General Circulation Model (TIEGCM), which solves the three-dimensional equations of continuity, momentum and energy for neutral and ion gases of the IT system. The TIEGCM version used here determines state variables globally on a uniform geographic grid of 1.25° in both latitude and longitude with 57 pressure levels at different altitudes (Dang et al., 2018; Qian et al., 2013; Wang et al., 2008). In addition to its inputs from LFM – convection velocity and electron precipitation flux and energy – TIEGCM uses the F10.7 index for solar radio flux to parameterize its extreme ultraviolet ionization.

The MIX solver couples TIEGCM and LFM on a uniform grid of 2° in MLAT and magnetic longitude and extends from the poles down to 46° MLAT. MIX takes conductance information from TIEGCM and the field-aligned current (FAC) distribution from LFM to solve a Poisson equation for the electric potential at the ionospheric height. The potential determines the ionospheric convection velocity in TIEGCM and a velocity boundary condition at LFM's low-altitude boundary after mapping it along equipotential, magnetic dipole field lines to the boundary (Merkin & Lyon 2010). The time interval between LFM-MIX-TIEGCM data exchanges is 5 s. TIEGCM's time step is set to be the same as the (5 s) exchange time.

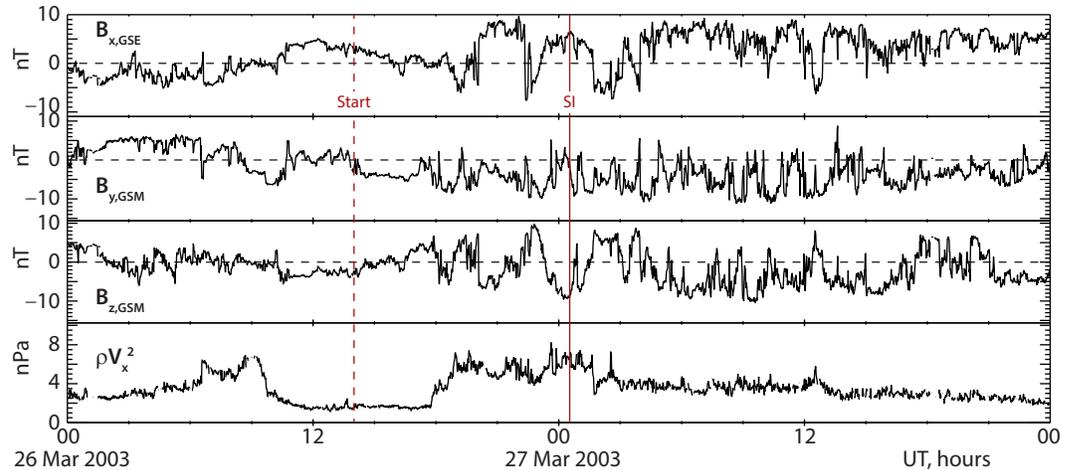

**Figure 2.** IMF and solar wind dynamic pressure from 00 UT on 26 March 2003 through 00 UT on 28 March 2003. The red dashed and solid vertical lines respectively mark the start of the SIR event at 1400 UT and its stream interface at 0038 UT (Jian et al., 2006), neither of which delineates a distinct transition in the plotted variables.

The low-altitude cusp finding algorithm developed by Zhang et al. (2013) from LFM state variables is used to define the cusp area. This simulated dynamic cusp is projected onto TIEGCM's grid where low-energy, direct-entry cusp electron precipitation is specified, also using LFM state variables. Direct-entry cusp precipitation is distinct from broadband, monoenergetic and diffuse precipitation which MIX also determines from LFM state variables and passes to TIEGCM (Zhang et al., 2015b). Low-energy electron cusp precipitation has been shown to be important in thermospheric heating in the cusp because it enhances the bottomside *F*-region Pedersen conductivity and thermospheric heating there (Clemmons et al., 2008; Zhang et al., 2012, 2015a; Deng et al., 2013; Brinkman et al., 2016). We use LFM's cusp finding algorithm to specify the TIEGCM cells in which Alfvénic heating is imposed (Section 2.4).

We used CMIT to investigate the thermospheric heating and upwelling that occurred as a stream interaction region (SIR) swept past Earth on 26-27 March 2003 (Figure 2). We chose an SIR event because variability in the solar wind and IMF accompanying such events is expected to produce intense Alfvénic activity in the cusp. During this event the CHAMP mean altitude (413 km) is comparable to the altitude where the distribution of density anomalies in Fig. 1 occur (Liu et al., 2005). Hemispheric asymmetries in the ionospheric conductivities are minimized (though not entirely absent) because the event is near spring equinox. No SIRs occurred in the 11 days prior to this event (Jian et al., 2006), so a less disturbed MIT system is expected at model start time, making the geospace response to the SIR reasonably distinct. The sampled IT environment begins to respond to the SIR around 00:00 UT on the 27th (described in Section 3) which is the middle of the simulated period. The first day of the event (until 1800 UT) exhibits weak-to-moderate activity in the IMF; the second day exhibits more intense and persistent variability in the IMF.

Using IMF data to drive LFM requires special treatment of a variable IMF $B_x$ in order to maintain $\nabla \cdot \mathbf{B} = 0$ in the interplanetary medium (Lyon et al., 2004). As described in Text S1 of the Supporting Information, we used the method of Wiltberger et al. (2000) to specify the variable IMF $B_x$ for this event. To reduce memory of initial conditions in simulating the SIR event, TIEGCM and LFM were separately preconditioned as described in Text S2 of the Supporting Information.

## 2.2 Parameterized Alfvén wave model

We consider energy deposition in the cusp ionosphere by low-frequency magnetohydrodynamic waves, presumed to be generated at high altitudes near the magnetopause by interplanetary variability and/or magnetopause dynamics associated with dayside reconnection and Kelvin-Helmholtz surface waves. In general, such disturbances stimulate slow and fast mode compressional waves and intermediate mode shear Alfvén waves. Slow mode waves encounter strong ion Landau damping in the magnetosphere where the ion to electron temperature ratio typically exceeds one. Consequently, very little slow mode power stimulated by high-altitude disturbances reaches the ionosphere. The group velocity of fast mode waves is nearly isotropic, so fast mode power falls off as $r^{-2}$ with distance $r$ from the source. The power carried by nondispersive, shear Alfvén waves reaches the ionosphere most efficiently because (nondispersive) shear Alfvén waves are guided practically loss-free along magnetic field lines, and their energy is magnetically focused by the converging flux tube. The field-aligned Poynting flux $S_{\parallel}$ of the Alfvén wave varies approximately with magnetic field intensity $B$ as $S_{\parallel}/B$. For these reasons, we consider energy deposition resulting only from shear mode Alfvén waves and neglect signals and ionospheric energy deposition resulting from slow and fast mode waves.

The linear model for collisional propagation and absorption of Alfvén waves in the ionosphere-thermosphere used in this study is described by Lotko & Zhang (2018). It neglects magnetic compressibility, ionospheric Hall currents and electron inertia, which, as discussed by Lotko and Zhang, is appropriate for angular wave frequencies ($\omega = 2\pi f$) and perpendicular wavenumbers $k_{\perp}$ satisfying $\mu_0 \, e^2 \, n_e/m_e \gg k_{\perp}^2 \gg \mu_0 \omega \sigma_H^2/\sigma_P$ for given Hall ($\sigma_H$) and Pedersen ($\sigma_P$) conductivities, the permeability of free space ($\mu_0$), and the electron charge ($e$), mass ($m_e$) and density ($n_e$). We also neglect the effects of neutral-gas inertia, mediated by ion-neutral friction, on Alfvén wave propagation and dissipation. These effects are negligible when $\omega \gg \nu_i \rho_i/\rho_n$ where $\rho_i$ and $\rho_n$ are the ion-species and neutral-gas mass densities and $\nu_i$ is the ion-neutral collision frequency (see Supporting Information [Text S3](#) and [Figure S2](#)).

The equations of Lotko & Zhang (2018) are:

$$\partial_t \mathbf{E}_{\perp} + \nu_P \mathbf{E}_{\perp} = V^2 \left( \nabla \times \mathbf{B} \right)_{\perp} \tag{1}$$

$$\partial_t \mathbf{B} = -\left( \nabla \times \mathbf{E} \right) \tag{2}$$

$$E_{\parallel} = j_{\parallel}/\sigma_0 = \eta \, \hat{\mathbf{b}} \cdot \left( \nabla \times \mathbf{B} \right) \tag{3}$$

$\mathbf{E}_{\perp}$ and $\mathbf{B}_{\perp}$ are the wave perpendicular electric and magnetic fields; $j_{\parallel}$ is the wave field-aligned current derived from $\mathbf{B}_{\perp}$; $\hat{\mathbf{b}}$ is the unit vector in the local magnetic-field direction. In (1) and (3) the Pedersen dissipation rate ($\nu_P = \sigma_P/\varepsilon$) and magnetic diffusivity ($\eta = 1/\sigma_0\mu_0$) are given in terms of the Pedersen and direct or parallel ($\sigma_0$) conductivities defined as

$$\sigma_P = \sum_s \frac{n_s q_s^2}{m_s} \frac{\nu_s}{\nu_s^2 + \Omega_s^2}, \tag{4}$$

$$\sigma_0 = \frac{n_e e^2}{m_e \nu_e} \, . \, . \tag{5}$$

The wave speed ($V = 1/(\mu_0\varepsilon)^{1/2} = c(\varepsilon_0/\varepsilon)^{1/2}$) in (2) depends on the speed of light in vacuum ($c$), the vacuum permittivity ($\varepsilon_0$) and the dielectric constant ($\varepsilon$) for Alfvén wave propagation including collisional effects

$$\varepsilon = \varepsilon_0 \left( 1 + \sum_s \frac{\omega_{ps}^2}{\nu_s^2 + \Omega_s^2} \right). \tag{6}$$

Plasma parameters in (4)-(6) include the number density ($n_s$), charge ($q_s$), mass ($m_s$), plasma frequency ($\omega_{ps}$) and collision frequency ($\nu_s$) of charged-particle species $s$ ($e$ for electrons, $i$ for ion species). The collision frequencies are taken from Schunk & Nagy (SN) (2009). For ions, $\nu_s$ includes ion-neutral collisions (SN, equation 4.146 and Table 4.4; Table 4.5); for electrons, it includes electron-ion (SN, equation 4.144) and electron-neutral collisions (SN, Table 4.6).

Our focus is on low-altitude collisional Alfvénic energy deposition (below 1000 km and mostly below 500 km), so we follow the method of Lotko & Zhang (2018) and simplify the analysis by treating the geomagnetic field as straight and uniform. The implications and accommodations of this treatment are described in [Text S4](#) of the Supporting Information.

At this juncture we neglect spatial gradients in MLT in the background parameters and in Alfvén waveforms, which is also assumed in deducing SSFACs from low-altitude satellite data (see below). The 2D form of (1)-(3) in coordinates parallel ($z$) and perpendicular ($x$) to the background magnetic field ($x$ corresponds to the magnetic meridional or MLAT direction, positive toward north) become, after combining (2) and (3),

$$\left(\frac{\partial}{\partial t} + \nu_P\right) E_x = -V^2 \frac{\partial B_y}{\partial z}, \tag{7}$$

$$\frac{\partial B_y}{\partial t} = -\frac{\partial E_x}{\partial z} + \frac{\partial}{\partial x}\left(\eta \frac{\partial B_y}{\partial x}\right). \tag{8}$$

We neglect perpendicular (MLAT) spatial variations in background parameters $\nu_P$, $V$ and $\eta$, which is valid when the length scales for such spatial variations are large compared to the perpendicular wavelength. Equations (7) and (8) are solved by harmonic solutions of the form

$$E_x(x,z,t) = \sum_{m=1}^{M}\sum_{n=1}^{N} E_{m,n}(z)\cos\left(k_m x + \theta_m\right)\cos\left(\omega_n t + \theta_n\right), \tag{9}$$

$$B_y(x,z,t) = \sum_{m=1}^{M}\sum_{n=1}^{N} B_{m,n}(z)\cos\left(k_m x + \theta_m\right)\cos\left(\omega_n t + \theta_n + \phi_{m,n}\right). \tag{10}$$

$E_{m,n}(z)$ and $B_{m,n}(z)$ are altitude-dependent Fourier amplitudes; $\phi_{m,n}(z)$ is the altitude-dependent phase difference between the wave electric and magnetic field for each mode; $k_m = 2\pi m/\lambda_{\max}$ is the wavenumber; $\omega_n = 2\pi n/\tau_{\max}$ is the angular frequency; and $\lambda_{\max}$ and $\tau_{\max}$ are the maximum meridional perpendicular wavelength (100 km) and wave period (20 sec) considered. Lacking knowledge of any correlations or relationships between the different modes, we assume they are uncorrelated with random phases $\theta_m$ and $\theta_n$ on the interval $[0, 2\pi]$.

The expansions include $M \times N = 400 \times 40$ modes. The following considerations were made in choosing the upper and lower cutoffs on perpendicular wavenumbers and frequencies.

–  Modes with perpendicular wavelengths $\lesssim 2$ km encounter strong Ohmic dissipation in the topside ionosphere (Lotko & Zhang, 2018), mainly due to electron-ion collisions. Ohmic dissipation heats electrons, with very little thermospheric heating. The expansions (9) and (10) include wavelengths as short as $\lambda_{\max}/M = 250$ m, but thermospheric heating and upwelling are relatively insensitive to inclusion of these short wavelength modes.

–  The longest wavelength included in (9) and (10) ($\lambda_{\max} = 100$ km) is an appropriate upper limit on the wavelength spectrum when modeling parameterized Alfvénic heating in CMIT when the TIEGCM cell size is 1.25° (140 km) and the MIX and LFM cell sizes are 220 km or larger when referenced to 100 km altitude.

–  Alfvén waves with frequencies comparable to or greater than the local proton gyrofrequency near the magnetopause are locally absorbed by ion-cyclotron resonant interactions (Stawarz et al., 2016). For this reason, we do not include modes with frequencies $> N/\tau_{\max} = 2$ Hz in (9) and (10) because they do not propagate from the near-magnetopause region to the low-altitude region of interest.

–  The lowest frequency mode ($f_{\min} = 1/\tau_{\max}$) in (9) and (10) is practically quasi-static. Including electric field variability with frequencies below $f_{\min}$ would augment the dc Joule heating rate in CMIT, which is treated in the Joule heating module of TIEGCM, as discussed in Section 2.4.

In solving (7) and (8), the harmonic dependence in x is first introduced which converts $\partial^2/\partial x^2$ in (8) to $-k_m^2$. The altitude dependent amplitudes $E_{m,n}(z)$ and $B_{m,n}(z)$ and phases $\phi_{m,n}(z)$, now parameterized by $k_m$, are then derived from the finite-difference time-domain (FDTD) numerical method described by Christ et al. (2002) with modes stimulated by a time-periodic driver with frequency $n/\tau_{\max}$ at altitude $z = 4500$ km and electric field amplitude $E_{m,n}(z_d)$. Setting $B_{m,n}(z_d) = E_{m,n}(z_d)/V(z_d)$ at the driver stimulates a downward propagating Alfvén wave. The computational domain extends from the Earth's surface ($z = 0$ km altitude) up to $z = 5000$ km. The FDTD solver with Mur absorbing boundary conditions (Mur, 1981) is run until its solution converges to the harmonic forms (9) and (10) (Lotko & Zhang, 2018).

We impose observational constraints on the FDTD solution by adjusting (through trial and error) the frequency and wavenumber dependence of $E_{m,n}(z_d)$ to produce an amplitude-frequency spectrum for the field-aligned current (FAC) similar to the average spectrum shown in Figure 3 of Rother et al. (2007), derived from CHAMP satellite measurements at an altitude of approximately 400 km. The procedure for imposing the constraints is as follows.

– As is typical when determining the FAC from satellite measurements of the magnetic field (Lühr et al., 1996), Rother et al. assume (i) the FAC is effectively sheet-like (also consistent with the 2D approximation used in (7) and (8)) and (ii) the curl of the magnetic field is equivalent to the along-track Doppler derivative normal to the sheet, i.e., $\partial/\partial x = \partial/V_s \partial t$. Here, x is the spatial coordinate normal to the sheet and the background magnetic field. $V_s \approx 7.6$ km/s is the component of the CHAMP satellite velocity normal to the sheet and the background magnetic field. The extent to which these observational assumptions are valid for the events reported by Rother et al. (2007) is not clear, but we simulate the procedure as follows:

– In (10), we set $x = V_s t$, differentiate the resulting expansion with respect to $V_s t$, and divide the result by $\mu_0$ to obtain the simulated, Doppler-derived FAC as a function of time.

– Sample time series of the Doppler-derived FAC along a simulated CHAMP satellite track may then be constructed and its Fast Fourier Transform calculated to obtain a sample amplitude-frequency spectrum. Choosing the n and m dependence of the modal amplitudes at the driver to be

$$E_{m,n}(z_d) = E_0 \Big/ n\sqrt{n^2 + 2.4m^2} \tag{11}$$

was found to approximate the spectral shape of the FACs reported by Rother et al. (2007). $E_0 = 500$ mV/m produced a peak FAC in the Doppler-derived time series comparable to the mean peak value of 350 $\mu$A/m$^2$ in Fig. 7 of Rother et al. (2007). The observed peak values range from 100 $\mu$A/m$^2$ (Rother's threshold value) to 2000 $\mu$A/m$^2$. $E_0$ was scaled up to 1500 mV/m to bracket a likely upper limit on thermospheric upwelling in the simulations with Alfvénic heating described in Sec. 3. Note that the actual electric field amplitude calculated from (9) is much less than $E_0$ owing to the mode number dependence in the denominator of (11).

Model time series and spectra obtained from the above procedure, using profiles for the background parameters specified in the next section, are described in the Supporting Information Text S5 and Figs. S3 and S4.

## 2.3 Specification of background parameters

FDTD solutions to (7) and (8) require specification of the altitude profiles (z dependence) for parameters $\sigma_P$, $\sigma_0$ and $\varepsilon$. We determined these parameters from equations (4)-(6) using the most recent International Reference Ionosphere (IRI-2016) (Bilitza 2018), the Navy Research Lab Mass Spectrometer and Incoherent Scatter radar (NRLMSISE-00, or MSIS) model (Picone et al., 2002) and an Earth-centered dipole model for the geomagnetic field. The IRI model uses the date to look up historical data of daily F10.7, a measure of solar activity by measuring the solar radio flux at 10.7 cm as a proxy of solar activity. The MSIS model similarly uses F10.7, as well as user-input for the 3-hour Ap, a proxy of the planetary magnetic field for which historical data exists.

Two different profiles for $\sigma_P$, $\sigma_0$ and $\varepsilon$ were computed at the average cusp location (given by its MLT-MLAT center) at 12:00 UT on 26 March 2003 and on 27 March 2003 in a baseline CMIT run without Alfvénic heating. The first day exhibits weak-to-moderate solar wind/IMF driving. More intense SIR activity in the interplanetary medium is very prominent during the second day (Fig. 2). The electron density computed by IRI is scaled such that the peak electron density (NmF2) matches the peak in the average electron density profile in CMIT in the cusp region for the same 24-hour period. This rescaling of the plasma density from IRI values to the mean CMIT values for each day of the simulated interval is a step toward self-consistency of the parameterized model and CMIT.

MSIS determines the thermosphere only up to 1000 km altitude. Above this altitude collisional effects are negligible and the collisionless Alfvén speed $(B/\sqrt{\mu_0 \sum_s m_s n_s})$ is used up to 2000 km. Above 2000 km altitude, the Alfvén speed is constant and equal to its value at 2000 km. The Pedersen conductivity is interpolated on a log-linear scale between 1000 and 2000 km altitude using MSIS-IRI at 1000 km and a specified small value (1.88x10$^{-12}$ S/m) at 2000 km altitude – the value in Lotko & Zhang (2018). The results are insensitive to this value as long as it is small. Above 2000 km we set $\sigma_P$ to its value at 2000 km. The parallel conductivity is interpolated on a log-linear scale from its computed value at 1000 km to a large value at the top of the simulation space such that η becomes negligible; $\sigma_0 = 1 \times 10^{10}$ S/m was chosen. At altitudes below 90 km, below the range of CMIT, the wave speed is constant and equal to 12000 km/s. The wave speed is actually larger than 12000 km/s in the lower thermosphere, approaches the speed of light in the lower atmosphere and is artificially limited here for numerical efficiency. This choice does not influence the results in the IT region of interest. The conductivity profiles are interpolated on a log-linear scale between computed values at 90 km and 0 km altitude where both $\sigma_P$ and $\sigma_0$ are set to $6 \times 10^{-12}$ S/m (Lysak 1999, Lotko & Zhang, 2018). The wave solution at *E*- and *F*-region altitudes is relatively insensitive to the value specified for conductivities at Earth's surface, provided it is small compared to the value above 90 km.

The altitude profiles of $\sigma_P$, $\sigma_0$ and wave speed $V$ used to compute FDTD solutions for the Alfvén wave fields are included in [Figure S5](#) of the Supporting Information.

### 2.4 Integrating Alfvén wave heating into CMIT

The Alfvén wave solution, observationally constrained as described in the previous section, is used to augment CMIT's Joule heating rate as a function of altitude. To this end, we use the fact that time scale for ion acceleration by the Alfvénic fluctuations (wave periods ≤ 20 sec) is short compared to the time scale for neutral-fluid acceleration ($v_{ni}^{-1} > 1$ hour). This disparity in time scale renders the Alfvén wave-induced quiver velocity of the neutral gas—mediated by ion-neutral friction—negligible compared to the perturbed ion velocity for the considered range of wave frequencies (0.5-2 Hz) and IT conditions described in Sec. 2.3 (see [Text S3](#) and [Fig. S2](#)). In this case, the effects of Alfvén wave acceleration on ion-neutral friction may be neglected in the neutral gas momentum equation, and the neutral gas may be considered stationary when evaluating the Alfvénic contribution to Joule heating (SI [Text S3](#)). Also making use of the inequalities $\omega^2, \omega v_i \ll \Omega_i^2$ that underlie the derivation of (1)-(3), the Alfvénic Joule heating rate $Q_{J,A}$ may be calculated as $\sigma_P \delta E_A^2$ (see [Text S7](#)).

Rather than specifying a particular realization of the Alfvénic fluctuations encountered by the neutral fluid as it traverses the cusp region, we characterize the fluctuations in terms of an ensemble average over all possible random phases in (9) to obtain the Alfvénic Joule heating rate in terms of the RMS amplitude of the fluctuations:

$$Q_{J,A}(z) = \sigma_P \left\langle E_x^2(x,z,t) \right\rangle_{x,t} = \sigma_P \frac{1}{\lambda_{max}} \int_0^{\lambda_{max}} dx \frac{1}{\tau_{max}} \int_0^{\tau_{max}} dt E_x^2(x,z,t) = \frac{\sigma_P}{4} \sum_{m,n} E_{m,n}^2(z) \equiv \sigma_P \delta E_{A,rms}^2. \quad (12)$$

The average profile for the root-mean-square (RMS) Alfvénic electric field, $\delta E_{A,rms}$, is precomputed twice by the FDTD Alfvén wave solver, at 12:00 UT on 26 March 2003 and on 27 March 2003, at the average cusp locations where the $\sigma_P$ profiles were determined from the TIEGCM-scaled, MSIS-IRI data as described in Sec. 2.3. These two representative $\delta E_{A,rms}$ profiles are taken to be constant in *x* and *t* throughout the corresponding day and over the instantaneous cusp area determined by LFM's cusp finding algorithm (Zhang et al., 2013). While the Alfvén wave power flowing into CMIT's cusp area is actually bursty in space and time, the thermosphere effectively integrates over the Alfvénic heating during its relatively slow transit through the cusp. Thus a space-time average for parameterized Alfvénic heating of the thermosphere is appropriate. The profile obtained for $\delta E_{A,rms}$ for day 2 of the event is shown in [Figure 3](#) along with that of a static electric field with the same value in the *E* region for comparison. The Alfvénic electric field is ≈ 50% larger at *F*-region altitudes than a static electric field of comparable intensity in the *E*-region ionosphere owing to the effect of the IAR. IAR modes at frequencies greater than ≈ 0.5 Hz are efficiently pumped by the driver, and they intensify upon being trapped by the strong opposing gradients in wave speed in the *F* region (cf. Figs. 2 and 3 of Lotko & Zhang, 2018).

The RMS electric field $\delta E_{A,rms}$ from the Alfvén wave model is integrated into the Joule heating module of CMIT by augmenting the

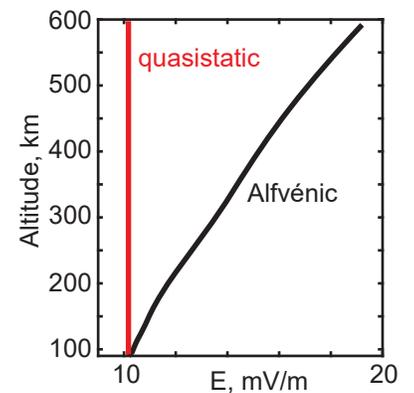

**Figure 3.** Altitude profile of the rms Alfvén wave electric field $\delta E_{A,rms}$ for day 2 of the event compared to that of a quasistatic electric field with the same amplitude in the *E* region.

Joule heating per unit mass for the neutral $\left( Q_J^{T_N} \right)$ and ion $\left( Q_J^{T_i} \right)$ gasses according to the formulae

$$Q_J^{T_N} = \frac{\sigma_P B^2}{\rho} \left[ C_J \left( \mathbf{v}_{E \times B} - \mathbf{v}_{n\perp} \right)^2 + \frac{\delta E_{A,rms}^2}{B^2} \right] \tag{13}$$

$$Q_J^{T_i} = \frac{\overline{m}_n}{\overline{m}_n + \overline{m}_i} Q_J^{T_N} \quad . \tag{14}$$

The Pedersen conductivity $\sigma_P$ in $Q_J^{T_N}$ is calculated by CMIT at each TIEGCM time step (5 seconds), $\rho$ is the thermospheric mass density from TIEGCM, $B$ is the geomagnetic field (the International Reference Geomagnetic Field is used in TIEGCM), $\mathbf{v}_{E \times B}$ and $\mathbf{v}_{n\perp}$ are, respectively, the ion $E \times B$ drift velocity and the neutral wind velocity perpendicular to $B$, both derived from CMIT, and $\overline{m}_n$ and $\overline{m}_i$ are the mean molecular mass of neutrals and ions, respectively. Equation (13) is the Joule heating rate in the reference frame of the neutral gas, which is equivalent to the frictional heating rate due to ion-neutral drag (see SI [Text S7]). For the reasons discussed in SI [Text S3], the neutral gas may be considered stationary when evaluating the Alfvénic contribution to Joule heating, i.e., the second term on the right side of (13)

The basic Joule heating module in TIEGCM includes the constant multiplicative factor ($C_J$) that augments the Joule heating specified by the first term in (13). This factor is intended to model the effects of quasistatic subgrid variability in the Joule heating (e.g., [Figure 1]) and effectively adds subgrid Joule heat where TIEGCM produces grid-resolved Joule heat. Although this simple scaling for the subgrid Joule heat is not likely to be accurate in detail, it improves TIEGCM's baseline prediction for the thermospheric mass density. The addition of the parameterized Alfvénic heating model in CMIT also captures subgrid heating processes, so we reduced the nominal value $C_J = 1.5$ in the default TIEGCM code to 1.35. A time-dependent, assimilative adjustment of $C_J$, as in the assimilation method of Sutton (2018), may further improve the fidelity of the baseline prediction.

Even though $\delta E_{A,rms}$ is constant during each day of the SIR event, the Alfvénic Joule heating rate evolves in CMIT as TIEGCM evolves because $Q_{J,A}(z)$ in (13) is calculated using TIEGCM's $\sigma_P$. This Alfvén wave coupling dynamically modulates Joule heating because the CMIT-determined $\sigma_P$ changes with dynamic variations in electron precipitation and ion-neutral chemistry, both of which are at play during the simulated SIR event. In theory and possibly in the future, both $\delta E_{A,rms}$ and $Q_{J,A}(z)$ could be regulated at the cadence of a CMIT time step by regulating the Alfvén wave driver amplitude in the FDTD solver using LFM's Alfvén wave Poynting flux module (Zhang et al., 2015b) and TIEGCM's IT dynamic state. The approach implemented here may be considered a proof-of-concept to determine the value of continuing such future developments.

## 3. Results and Discussion

We first ran CMIT for 48 hours for the solar and interplanetary conditions during 26-27 March 2003 without the Alfvénic heating term in equation (13). This run provides a baseline for comparing changes when Alfvénic heating is included in CMIT.

A one-orbit moving mean of the CHAMP data and CMIT results ([Figure 4]) was evaluated to gain an overview of cumulative effects and hour-scale trends in the thermospheric response to the SIR event. The IMF activity and solar wind driving are weak to moderate during the first 24 hours (26 March 2003) of the event, and the air density recorded along the CHAMP orbit increases modestly during the day, from about $1.6 \times 10^{-12}$ kg/m³ to $2.3 \times 10^{-12}$ kg/m³. CMIT under-predicts the density by as much as 15% during the first day. Two curves are plotted for CMIT in [Fig. 4] to bracket its prediction. The curve labeled CMIT is the baseline estimate. The curve labeled CMIT with Alfvénic heating is calculated for an RMS wave electric field ($\delta E_{A,rms}$) that yields a peak Alfvénic FAC of 1 mA/m² ($E_0 = 1500$ mV/m in equation 11). The CMIT estimate for the orbit-average mass density of the neutral gas lies in the gray area between the two curves. The MSIS prediction is also plotted in [Fig. 4] for comparison.

Around 18:00 UT on 26 March, both the IMF activity and solar wind driving pick-up (cf. Fig. 2). The latency in the response of the thermosphere along this particular orbit is about 6 hours, after which the thermospheric density observed by CHAMP, simulated by CMIT and estimated by MSIS all begin a comparatively steep increase shortly after 00:00 UT on March 27 (cf. Fig. 4). All three estimates of the air density exhibit a relative maximum just after 18:00 UT on 27 March. CHAMP registers a density of $3.9\times10^{-12}$ kg/m³ at this time; the corresponding MSIS estimate is near $3.0\times10^{-12}$ kg/m³ while the CMIT prediction ranges from 3.6 to $3.8\times10^{-12}$ kg/m³. CMIT clearly tracks the CHAMP density more closely during the second day of the event. As expected the CMIT prediction is more dynamic than MSIS – a consequence of the minute-cadence interplanetary dynamics driving CMIT vs. the daily F10.7 and 3-hour Ap indices parameterizing MSIS.

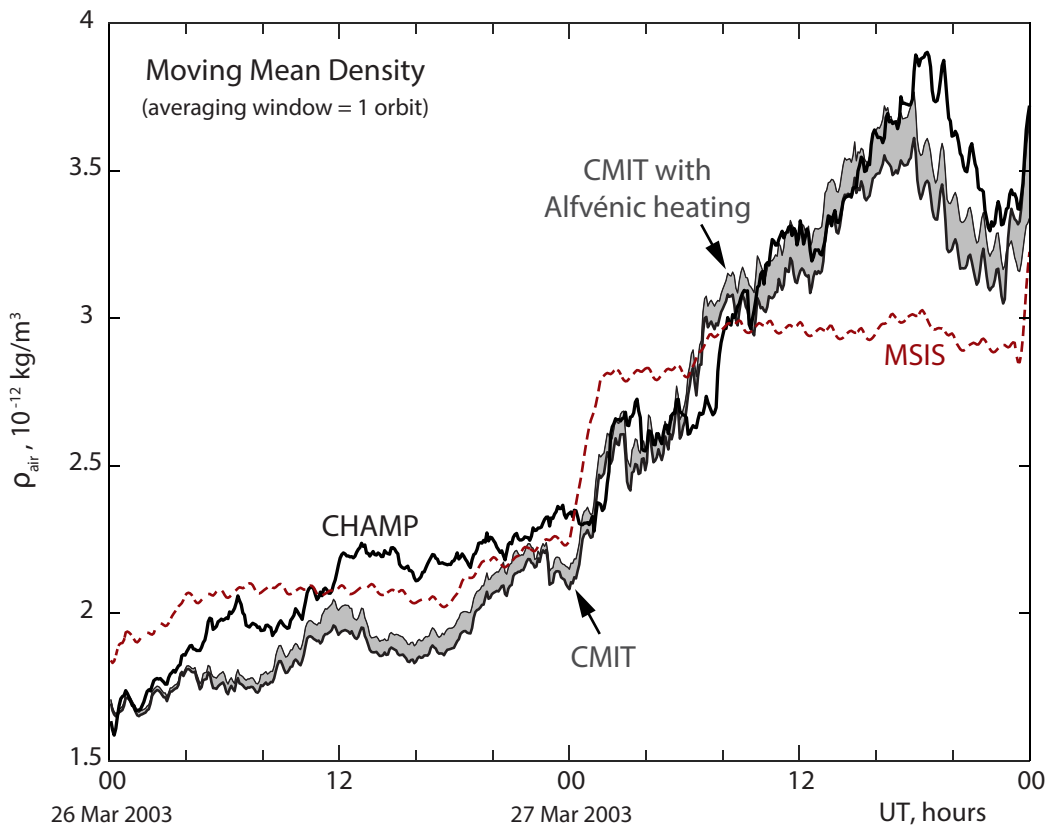

**Figure 4.** One-orbit moving mean of CHAMP-derived thermospheric mass density for 26-27 March 2003, CMIT simulated thermospheric density along the CHAMP orbit, with and without Alfvénic heating, and the MSIS prediction of the density along the CHAMP orbit. The curve labeled CMIT is the baseline run. CMIT with Alfvénic heating is the run with average peak FAC of 1 mA/m². Gray area between them indicates the expected range for CMIT density enhancement with Alfvénic heating effects.

The increase in air density due to Alfvénic heating relative to baseline is larger during the second day of the event than during the first day, with a maximum difference of $0.08\times10^{-12}$ kg/m³ during the first day, $0.15\times10^{-12}$ kg/m³ during the second day. However, the maximum difference over baseline is $\approx 4\%$ on both days because the baseline is greater on the second day. The larger increase in density from Alfvénic heating on the second day is due in part to an increase in TIEGCM's Pedersen conductivity.

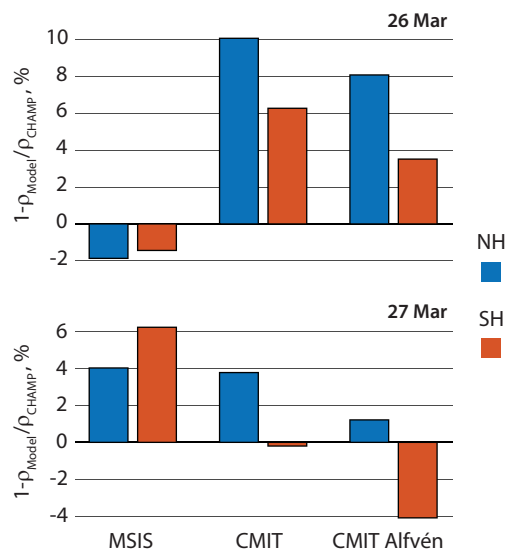

**Figure 5.** Comparison of daily mean error (in %) in model predictions relative to CHAMP's air density for MSIS, CMIT (baseline) and CMIT with Alfvénic heating for each hemisphere (NH, SH) and event day (26, 27 March).

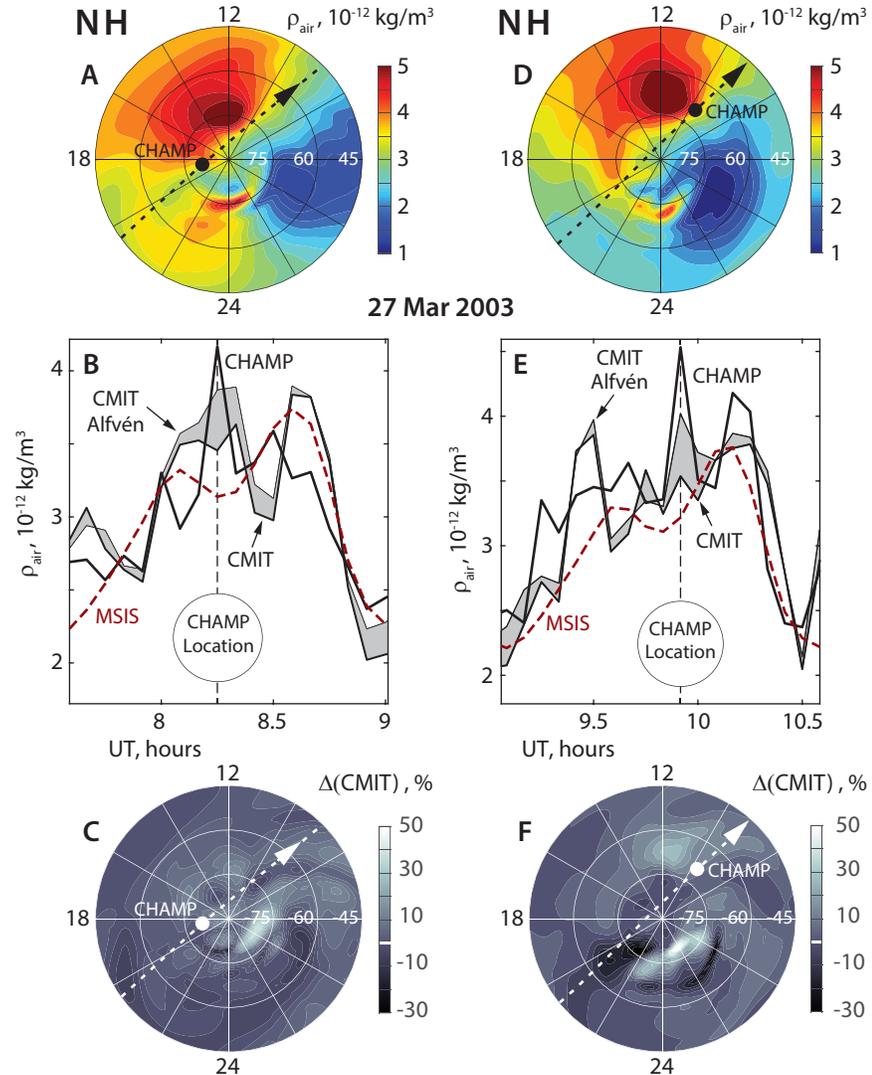

**Figure 6.** Air density along consecutive CHAMP orbits in the northern hemisphere (NH) on 27 March 2003 when CHAMP passes through the edges of high-latitude density enhancements. **A:** Simulated air density (color) at the CHAMP altitude (≈ 413 km) vs MLT and MLAT at the UT corresponding to "CHAMP Location" in panel B. Black circle indicates location of CHAMP at that UT. **B:** Comparison of instantaneously observed, simulated and MSIS air density vs. UT. CMIT curve is the baseline run. Curve labeled "CMIT Alfvén" is a run with a peak Alfvénic FAC of 1 mA/m². As in Fig. 4 gray area indicates expected range of CMIT predictions with Alfvénic heating. **C:** % difference in CMIT air density with and without Alfvénic heating. $\Delta(\mathrm{CMIT}) = \rho_{Alfvén}/\rho_{baseline} - 1$. Panels **D**, **E** and **F** in same format.

To quantify the average daily cumulative error in model predictions, we calculated the instantaneous error $(1 - \rho_{\mathrm{Model}}/\rho_{\mathrm{CHAMP}})$ along the CHAMP orbit and then determined the daily average error for each hemisphere (Figure 5). By this measure, MSIS performs better than CMIT for weak-to-moderate activity (26 March), with Alfvénic heating decreasing CMIT's relative error by ≈ 2% in both hemispheres. For higher activity (27 March), CMIT performs better than MSIS. Including Alfvénic heating in CMIT decreases its error by ≈ 3% in the NH on 27 March, but it increases the error by 4% in the SH. We attribute the reduction in error when Alfvénic heating is included, in part, to the increased Joule heating resulting from the increase in CMIT's Pedersen conductivity. The increase in error in the SH during 27 March may be due in part to inconsistencies in CMIT's treatment of the geomagnetic field, which is based on the International Geomagnetic Reference Field in TIEGCM and on a Earth-centered dipole field in the LFM and MIX solvers. TIEGCM includes the intrinsic NH/SH asymmetries of the IGRF, but LFM and MIX drive TIEGCM with inputs from the magnetosphere that ignore this asymmetry.

We also compared the instantaneous air density recorded on CHAMP with CMIT and MSIS predictions. The left panel of Figure 6 shows CMIT diagnostics for the air density on 27 March 2003 in the northern hemisphere (NH) at the CHAMP altitude (413 km). At 0815 UT CHAMP traverses the edge of a dayside density enhancement in the CMIT thermosphere including the effects of cusp-region Alfvénic heating (Fig. 6A). The point of closest approach at 0815 UT is indicated by the black circle located at 18.7 MLT and 81.0° MLAT. Fig. 6B compares the instantaneous air densities from CHAMP measurements, CMIT predictions and MSIS estimates along the CHAMP orbit vs UT. As in Fig. 4, the curve labeled CMIT is for the baseline run, while the curve labeled CMIT with Alfvénic heating is the CMIT run with a peak Alfvénic FAC of 1 mA/m². The gray area between these two curves indicates the expected range of CMIT predictions with Alfvénic heating. Fig. 6C is a NH map of the % difference in CMIT's predictions with and without Alfvénic heating: $\Delta(\mathrm{CMIT}) = \rho_{Alfvén}/\rho_{baseline} - 1$. The maximum density difference in Fig. 6B is about 12%, but the main cusp density enhancement (not traversed by CHAMP) exhibits a difference closer to 20% in Fig. 6C. Changes in thermospheric density resulting from Alfvénic heating are not localized to the cusp or the dayside. The largest difference occurs between about 0200 and 0600 MLT near 80° MLAT. The effects of co-rotation and neutral winds evidently redistribute the additional cusp density enhancement produced by Alfvénic heating.

The right panels of Fig. 6 continue the analysis in the same format for the subsequent CHAMP crossing of the NH. CHAMP traverses the edge of the density enhancement in CMIT at 9.67 MLT and -69.5° MLAT at 0955 UT (black circle in Fig. 6D). The comparison between

CHAMP measurements, CMIT prediction and MSIS estimate along the orbit (Fig. 6E) shows that CMIT predicts the location of CHAMP's peak density in both hemispheres, and its accuracy improves with Alfvénic heating included. This feature is missed entirely by MSIS, which exhibits a relative minimum where both the CHAMP and MSIS densities peak. The maximum density difference for CMIT with and without Alfvénic heating is about 14% in Fig. 6E. As for the previous NH crossing, the main cusp density enhancement (which CHAMP does not traverse) exhibits a larger difference closer to 30% in Fig. 6F, and the redistribution and modification of the Alfvénic-enhanced density is prominent throughout the nightside polar region.

## 4. Conclusions

- CMIT tracks orbit-averaged air density at CHAMP altitudes more closely than MSIS during intervals of solar wind and IMF activity characteristic of an SIR event (Fig. 4).

When comparing the IT response during the less active, first day of the simulated SIR event to the response during the more active second day, we find that CMIT replicates CHAMP-derived density measurements better than MSIS during the more active period. This finding is not surprising because the proxy-governed nature of MSIS relies on inputs of average parameters over longer periods (multi-hour to days) and is not as responsive to interplanetary variability as CMIT when driven by one-minute solar wind and IMF data. The MIT coupling intrinsic to CMIT allows effects such as soft electron precipitation, electric field variability and Alfvénic heating to modify the Pedersen conductivity, Joule heating rate and, therefore, air temperature and density in the cusp region on faster time scales than MSIS.

- Alfvénic heating modestly improves CMIT's daily mean prediction of orbit-averaged air density relative to CHAMP measurements (Fig. 5) and produces 20-30% regional enhancements in the high-latitude thermospheric density relative to a CMIT simulation without Alfvénic heating (Fig. 6 bottom panels).

Inclusion of Alfvénic heating effects in CMIT decreases its daily mean error by 2-3% in orbit-averaged air density relative to measurements along a particular CHAMP orbit, except in the southern hemisphere during the active second day of the SIR event when the error increased by 4%. This discrepancy may arise in part from the nonconforming geomagnetic field models used in CMIT's three component models, which treat geomagnetic hemispheric asymmetry differently. Enhancements of 20-30% in air density are seen throughout the high-altitude region (not just the cusp region) relative to a baseline CMIT run. This finding suggests that, during the course of the two-day long simulation, co-rotation and neutral winds redistribute the additional cusp density enhancements produced by Alfvénic heating.

- Alfvénic heating produces significant thermospheric density enhancements at CHAMP altitudes in and near the cusp during times of intense Alfvénic Poynting flux (Fig. 6 middle panels).

Alfvénic thermospheric heating is especially effective when magnetopause variability induced by solar wind and IMF activity stimulates intense Alfvénic Poynting fluxes flowing into the low-altitude cusp. The resulting Alfvén wave energy deposition in the ionospheric Alfvén resonator adds to the important effects of soft electron precipitation and quasistatic electric field variability on $F$-region Joule heating and thermospheric upwelling previously reported in the literature. These relatively fast dynamic effects operating in concert may explain the anomalous thermospheric densities CHAMP recorded relative to the MSIS empirical model (Liu et al., 2005). A detailed analysis showed that MSIS completely misses the air density enhancements registered by both CHAMP and CMIT along near cusp orbits; and CMIT with Alfvénic heating substantially improves CMIT's baseline prediction of the enhancements (by 10-15%). During this SIR event, CHAMP did not actually traverse the central cusp region where CMIT's density enhancements with Alfvénic heating are even larger (20-30%) than without it.

## Acknowledgments

CHAMP and MSIS data were provided by Eric Sutton. CHAMP data are archived at http://tinyurl.com/den-sitysets. IMF and Solar wind data were provided by J.H. King, N. Papatashvilli at AdnetSystems, NASA GSFC and CDAWeb (http://omniweb.gsfc.nasa.gov/). We thank Roger Varney for sharing his Fortran code for collision frequencies, Jing Liu and Wenbin Wang for helpful insights into TIEGCM and thermospheric dynamics, and Wenbin for feedback on the manuscript. All code needed to produce the plots shown including CHAMP, MSIS, and CMIT data is available publicly: http://doi.org/10.5281/zenodo.3727080. We acknowledge support by the National Center for Atmospheric Research, a major facility sponsored by the National Science Foundation under Cooperative Agreement No. 1852977 and the Thayer School of Engineering at Dartmouth College. Computing resources were provided by NCAR's Computational and Information Systems Laboratory (CISL). BH was supported by NASA NH Space Grant NNX15AH79; WL by NASA Grant 80NSSC19K0071, and WL and KP by NSF awards 1739188, 1522133.

**Alfvénic Thermospheric Upwelling in a Global Geospace Model**

Benjamin Hogan[1,2,†], William Lotko[1,2] and Kevin Pham[2]

[1]Thayer School of Engineering, Dartmouth College, Hanover, NH, USA
[2]High Altitude Observatory, National Center for Atmospheric Research, Boulder, CO, USA
[†]Now at Laboratory for Atmospheric and Space Physics, University of Colorado Boulder, Boulder, CO, USA
   and Department of Aerospace Sciences, University of Colorado Boulder, Boulder, CO, USA

## Contents of Supporting Information

## Introduction

This Supporting Information includes the following sections and figures:

Text S1. CMIT treatment of IMF $B_x$ and Figure S1 comparing $B_x$ from the OMNI data set with the constrained $B_x$ used in the SIR simulation.

Text S2. CMIT preconditioning.

Text S3. Validity of neglecting neutral gas inertia in collisional Alfvén wave dynamics and Figure S2 showing the domain of validity for the model ionosphere and thermosphere.

Text S4. Implications of treating the cusp-region geomagnetic magnetic field as straight and uniform and adjustments of model parameters to accommodate this simplification.

Text S5. Illustrative figures of time series of Alfvénic events constrained by CHAMP data (Figure S3) and amplitude-frequency spectra of modeled and observed field-aligned current events (Figure S4).

Text S6. Altitudes profiles for $V$, $\sigma_P$ and $\sigma_0$ (Figure S5) used to compute FDTD solutions.

Text S7. Notes on Joule and frictional heating.

## Text S1. CMIT Treatment of IMF $B_x$

Single point measurements acquired by a solar wind monitor are typically linearly advected in the direction of the solar wind flow, which is very nearly the GSE $x$ direction. Consequently, the time variability in IMF $B_x$ implies $\partial B_x/\partial x \neq 0$. Since information on the $y$ and $z$ dependence of IMF $B_y$ and $B_z$ is not available from single-point upstream measurements, the solenoidal condition, $\nabla \cdot \mathbf{B} = 0$, cannot be maintained without introducing additional constraints on the IMF. We follow the procedure of Wiltberger et al. (2000)[1] and assume the variability in IMF $B_x$ can be represented as

$$B_x(t) = a + bB_y(t) + cB_z(t).$$ (S1.1)

The coefficients $a$, $b$, and $c$ are determined by a multiple linear regression fit to all one-minute IMF data samples for the two-day SIR event (60 samples per hour × 48 hours). The fit establishes a plane oriented at a fixed angle relative to the GSE $x$ direction with normal direction

$$\vec{\mathbf{n}} = \left( \frac{1}{a}\hat{x} - \frac{b}{a}\hat{y} - \frac{c}{a}\hat{z} \right)$$ (S1.2)

along which the normal magnetic field is constant. This constraint allows a time-dependent $B_x$ to be introduced into the simulation that maintains $\nabla \cdot \mathbf{B} = 0$. For the 26-27 March 2003 event, the fitting procedure gives: $a = Bx_{fit}(0) = 0.99528304$; $b = By_{coef} = -0.24258055$; $c = Bz_{coef} = -0.56645968$. Figure S1 compares $B_x$ from the OMNI dataset with the value used in the CMIT simulation and that calculated from S1.1.

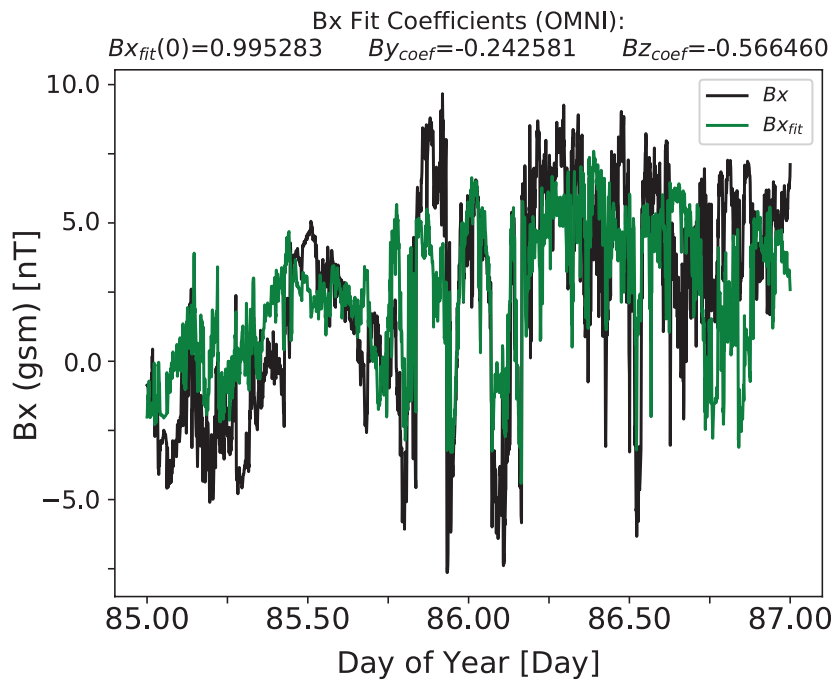

**Figure S1.** Comparison of measured IMF $B_x$ from the OMNI dataset with the values resulting from the multiple linear regression fit for the 26-27 March 2003.

## Text S2. CMIT preconditioning

CMIT is preconditioned for runs with real-time solar wind inputs as follows. The magnetosphere portion of CMIT, LFM, is typically preconditioned for four hours and fifty minutes before the start of the simulation time (in this study 00 UT March 26 2003).

1. LFM is first initialized with an Earth-centered dipole magnetic field permeating the entire simulation domain, populated by a low density fluid ($\approx 0.1$ cm$^{-3}$) composed of protons and alpha particles with number density ratio $n_{He}/n_p = 0.04$.

2. For the first fifty minutes of the LFM preconditioning, the upstream boundary conditions on the IMF and solar wind are specified with a 0 nT IMF, solar wind velocity ($V_x, V_y, V_z$) = ($-400,0,0$) km/s, number density $n_{SW} = 5$ cm$^{-3}$ and sound speed $c_{SW} = 40$km/s. As the resulting solar wind dynamic pressure interacts with the dipole magnetic field, it compresses the magnetic cavity on the dayside and stretches the nightside field to form the expected magnetospheric shape. This first phase of initialization prevents the zero-IMF solar wind front from advecting onto a grid cells containing dipole magnetic field lines so the resulting magnetosphere is in the initial low-density, near-vacuum state.

3. At the end of this 50-minute period, the IMF at the upstream boundary is set to ($B_x, B_y, B_z$) = (0,0,$-5$) nT for the next two hours with the same solar wind parameters. When the southward IMF contacts the magnetopause, the resulting magnetic reconnection opens the magnetosphere and allows solar wind fluid to enter the magnetosphere.

4. At 2 hours, 50 minutes into the preconditioning, IMF $B_z$ is changed to $+5$ nT, which produces a quiet state and serves to expand and fully populate the magnetosphere with fluid from the solar wind.

5. The conductances required by MIX during the standalone LFM-MIX preconditioning are calculated from empirical models for precipitation[2] and EUV-induced[3] ionization. These conductances are obtained from TIEGCM after the models are coupled starting at 00 UT 26 March 2003.

CMIT's ionosphere-thermosphere model (TIEGCM) is preconditioned for five days (starting at 00 UT 21 March 2003) before starting the CMIT simulation interval at 00 UT 26 March 2003. For the preconditioning, standalone TIEGCM uses three empirical models as inputs, two of which (high-latitude electric field and electron precipitation models) are changed at 00 UT 26 March 2003 to LMF-MIX inputs. The high-latitude electric field and resulting ionospheric drift velocity during preconditioning are derived from the Kp-parameterized, Heelis empirical model[4]. The extreme ultraviolet (EUV) solar irradiance required by TIEGCM is obtained from the F10.7-parameterized, empirical EUVAC model[5]. TIEGCM uses its default Kp-parameterized, empirical auroral precipitation model during the five-day initialization. TIEGCM retains a history file of Kp and F10.7 values by date, which regulate the empirical models during the five-day preconditioning. EUVAC continues to be regulated by the historical F10.7 during 00 UT 26 March 2003 to 00 UT 28 March 2003.

LFM-MIX and TIEGCM initializations end at 00 UT 26 March 2003, when they are coupled and run as the CMIT model. In CMIT mode, LFM-MIX provides high-latitude convection and electron precipitation inputs to TIEGCM, while TIEGCM provides the Pedersen and Hall conductances required by the MIX potential solver. As described in Sec. 2.4, the pre-computed electric field in the Alfvén wave subgrid model is also included in TIEGCM at this time, first with the precomputed value for 26 March 2003, then with the precomputed value for 27 March 2003. When LFM-MIX and TIEGCM are first coupled, immediately after the preconditioning as standalone models, an impulsive change of state occurs in the simulated geospace environment due to discontinuous changes in the conductances used by MIX and the high-latitude convection and precipitation in TIEGCM as well as the addition of Alfvénic Joule heating to TIEGCM. To facilitate relaxation of the impulsive change before the onset of magnetic activity in the SIR event at approximately 1800 UT on 26 March 2003 (cf. Fig. 2) we chose to start running CMIT 18 hours earlier at 00 UT 26 March 2003. This 18-hour buffer period serves to facilitate relaxation of the impulsive change of state in CMIT.

Preconditioning intervals longer than five days for TIEGCM yield modestly different initial states at the end of the preconditioning interval, e.g., differences in neutral density at CHAMP altitudes for twenty-day preconditioning differ from values obtained for five-day preconditioning by less than 20%. Our controlled simulation experiments are concerned primarily with differences in CMIT predictions with and without parameterized Alfvénic heating, so the results of interest are relatively insensitive to moderate differences in IT states obtained from different preconditioning intervals. The buffer interval starting at 00 UT on 26 March 2003 before the onset of intense IMF activity of the SIR event 18 hours later further moderates any sensitivities to preconditioning after 18 UT on 26 March 2003.

### Text S3. Validity of neglecting neutral gas inertia in collisional Alfvén wave dynamics

Neutral-gas inertia, mediated by ion-neutral collisions, has been shown to modify Alfvén wave characteristics at sufficiently low wave frequencies and high ion-neutral collision frequencies.[6,7] We derive the conditions of validity that allow this effect to be neglected for the model ionosphere and Alfvén wave spectrum considered in the paper.

The momentum equation for a cold neutral gas with velocity $\mathbf{v}_n$, mass $m_n$ and number density $n_n$ interacting with ion species $i$ with velocity $\mathbf{v}_i$, mass $m_i$ and number density $n_i$ via friction with collision frequency $\nu_{ni}$ is

$$n_n m_m \frac{\partial \mathbf{v}_n}{\partial t} = -n_n m_n \nu_{ni} \left( \mathbf{v}_n - \mathbf{v}_i \right) = n_i m_i \nu_{in} \left( \mathbf{v}_i - \mathbf{v}_n \right). \tag{S3.1}$$

The last equality follows from momentum conservation during collisions: $m_n n_n \nu_{ni} = m_i n_i \nu_{in}$. Momentum transfer from electrons is less by a factor of order $m_e \nu_{en} / m_i \nu_{in}$ and has been neglected.

Summing (S3.1) over $n$, assuming all neutral species have the same wind velocity $\mathbf{u}_n$, gives

$$\rho_n \frac{\partial \mathbf{u}_n}{\partial t} = -\sum_n n_n m_n \nu_{ni} \left( \mathbf{u}_n - \mathbf{v}_i \right) = n_i m_i \nu_i \left( \mathbf{v}_i - \mathbf{u}_n \right) \tag{S3.2}$$

where $\rho_n \equiv \sum_n n_n m_n$ and $\nu_i \equiv \sum_n \nu_{in}$.

Define $\delta_i \equiv n_i m_i / \rho_n$, assume time-harmonic variation, and Fourier Transform (S3.2) with respect to time to obtain the following relationship between Fourier amplitudes (designated as $\tilde{\mathbf{u}}_n$ and $\tilde{\mathbf{v}}_i$):

$$\tilde{\mathbf{u}}_n = \frac{\delta_i \nu_i}{-i\omega + \delta_i \nu_i} \tilde{\mathbf{v}}_i. \tag{S3.3}$$

---

The modulus of (S3.3) is

$$|\tilde{u}_n| = \frac{\delta_i \nu_i}{\sqrt{\omega^2 + \delta_i^2 \nu_i^2}} |\tilde{v}_i|.$$

(S3.4)

Lysak's (1999)[8] wave equations, from which equations (1)-(3) of Sec. 2.2 are derived, assume the neutrals are effectively stationary $(u_n \ll v_i)$. Equation (S3.4) shows that this assumption is valid when $\omega \gg \delta_i \nu_i$. In this case, the inertia of the neutral gas impedes its response to rapid oscillations in the ion velocity. Then

$$m_i n_i \frac{\partial \mathbf{v}_i}{\partial t} = -m_i n_i \nu_i (\mathbf{v}_i - \mathbf{u}_n) + e n_i (\mathbf{E} + \mathbf{v}_i \times \mathbf{B})$$

$$\approx -m_i n_i \nu_i \mathbf{v}_i + e n_i (\mathbf{E} + \mathbf{v}_i \times \mathbf{B}).$$

(S3.5)

This approximation is well-satisfied (Fig. S2) for the model ionosphere-thermosphere (described in SI Text S6 and Figure S5) and range of wave frequencies treated in the paper, especially in the lower ionosphere.

Collisional momentum exchange is conserved between ion species and the neutral gas, which is enslaved to the (Alfvénic) oscillating ion gas when $\omega \gg \delta_i \nu_i$: $\rho_n \partial \mathbf{u}_n / \partial t \cong n_i m_i \nu_i \mathbf{v}_i$. From this relation and (S3.4) with $\omega \gg \delta_i \nu_i$, we estimate the magnitude of the neutral-gas "quiver" velocity induced by Alfvén-wave ion oscillations as $u_n \sim (\delta_i \nu_i / \omega) v_i$. The maximum Alfvén wave electric field in the cusp is ~ 100 mV/m (cf. Fig. S3) with an implied maximum E×B velocity of ~ 2 km/s. Using the minimum frequency of 0.05 Hz, the factor $\delta_i \nu_i / \omega$ from Fig. S2, is 0.04, $10^{-4}$, and $10^{-6}$ at altitudes of 600 km, 300 km and 150 km, respectively. The maximum quiver velocity in the neutral wind is then ~ 80 m/s near the exobase, ~ 4 m/s in the $F$ region, and 0.04 m/s in the $E$ region. The average ion velocity encountered by the neutral gas in its transit through the cusp is about 50x less for the example in Fig. S3, and it is less by an additional 10x for Alfvén waves with frequencies (> 0.5 Hz), which are most effective in augmenting the Joule heating rate and thermospheric upwelling at $F$-region altitudes.[9] From these estimates, we conclude that Alfvénic ion forcing of neutral-gas momentum in the cusp is negligible compared to other thermospheric forcings (e.g., frictional quasi-static convection, pressure gradients, advection, etc.).

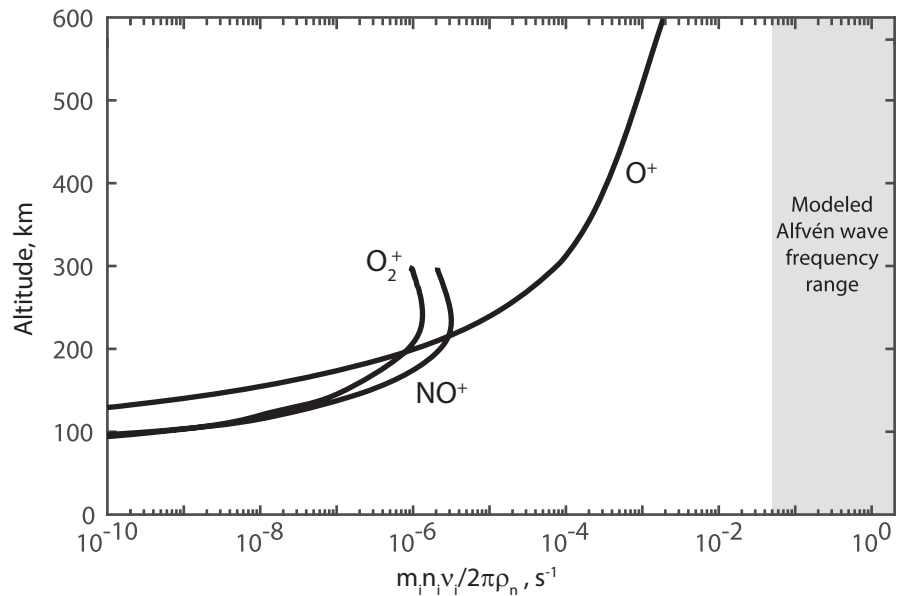

**Figure S2.** Altitude pro-file of $m_i n_i \nu_i / 2\pi \rho_n$ for O+, NO+, and O$_2$+ (domi-nant ion composition) vs altitude and comparison with the range of Alfvén wave frequencies (gray region, 0.05 Hz to 2 Hz) considered in the paper. Ion and neutral densities and ion-neutral collision frequencies are derived from IRI and MSIS mod-els evaluated in the cusp region for 27 March 2003 (see SI Text S6).

### Text S4. Implications of treating the geomagnetic field as straight and uniform

The actual spatial variation in the background magnetic field[10] introduces the following effects for an Alfvén wave propagating from 1000 km to 90 km altitude at the nominal magnetic latitude (75°) of the cusp: (i) The magnetic field increases by a factor of 1.5, which increases the speed of wave propagation; (ii) the flux tube convergence reduces the perpendicular length scale of the wave structure by a factor of 0.82; and (iii) the field line curvature shifts the geographic location of the magnetic foot point to higher latitude by about 1°. Effect (i) is included in the calculation of the dielectric constant ε in equation (6) and the wave speed $V$. Effect (ii) tends to reduce the Joule heating rate and increase the Ohmic heating rate, in going from high to low altitude and is modeled as described below. The small effect (iii) influences the interpretation of the geomagnetic latitude of wave energy deposition.

To include effect (ii) in the Alfvén wave solution, we follow Lotko & Zhang (2018)[9] and note that the radial variation of the square of the effective perpendicular wavenumber $k^2(r)$ due to magnetic focusing is $k^2(r) = k_{400}^2 (r_{400}/r)^3 = (2\pi/\lambda_{400})^2 (r_{400}/r)^3$ where $k_{400}$ is the wavenumber specified at 400 km altitude, $r_{400} \equiv 1$ R$_E$ + 400 km = 6800 km and $r = 1$ R$_E$ + $z$ is the radial distance at altitude $z$. The altitude dependence of the wave solution due to the altitude dependence of the wavenumber may be absorbed into an effective magnetic diffusivity defined as $\eta_{eff}(z) \equiv [r_{400}/(R_E + z)]^3/\mu_0 \sigma_0(z)$. With this definition, the solutions for the Alfvén wave electric field are parameterized by the perpendicular wavelength at 400 km altitude.

### Text S5. Alfvénic time series and observed and model spectra for FACs

Choosing modal amplitudes of the Alfvén wave electric field at the driver to be

$$E_{m,n}(z_d) = \frac{500 \text{ mV/m}}{n\sqrt{n^2 + 2.4m^2}} \tag{S4.1}$$

yields a peak amplitude for the FAC of approximately 350 µA/m² at 400 km altitude, the mean peak value in Fig.7 of Rother et al. (2007)[11]. A sample time series for the Doppler-derived FAC is shown in Fig. S3, calculated from the Doppler derivative $\partial/V_s\partial t$ of equation (10). Time series for the Alfvén wave electric and magnetic fields, curl-derived FAC (spatial rather than satellite Doppler derivative) and Joule heating rates are also shown in [Fig. S3](). The amplitude-frequency spectrum from the octave averaged FFT of the FAC in this time series is shown in [Fig. S4](), superposed on the average spectrum from Fig. 3 of Rother et al. (2007)[11]. The shape of the model spectrum approximates that of observed average spectrum reasonably well at low-frequencies ($f <$ 5 Hz), but it rolls-off more gradually above 5 Hz. The model spectrum at frequencies > 5 Hz is due to Doppler-shift of high wavenumber ($k$) modes for which the frequency in the spacecraft frame is $f = kV_s$, with $V_s \approx 7.6$ km/s. Ohmic rather than Joule dissipation dominates the absorption in this range (cf. Figs. 2 and 3 of reference[9]). Our results for thermospheric Joule heating are relatively insensitive to this high-$k$ dissipation.

---

[10] The following analysis is for an Earth-centered dipole field with an equatorial surface value of $3.1 \times 10^5$ T.

[11] Rother, M., Schlegel, K., & Lühr, H. (2007). CHAMP observation of intense kilometer-scale field-aligned currents, evidence for an ionospheric Alfvén resonator. *Annales de Geophysique, 25*(7), 1603–1615. https://doi.org/10.5194/angeo-25-1603-2007.

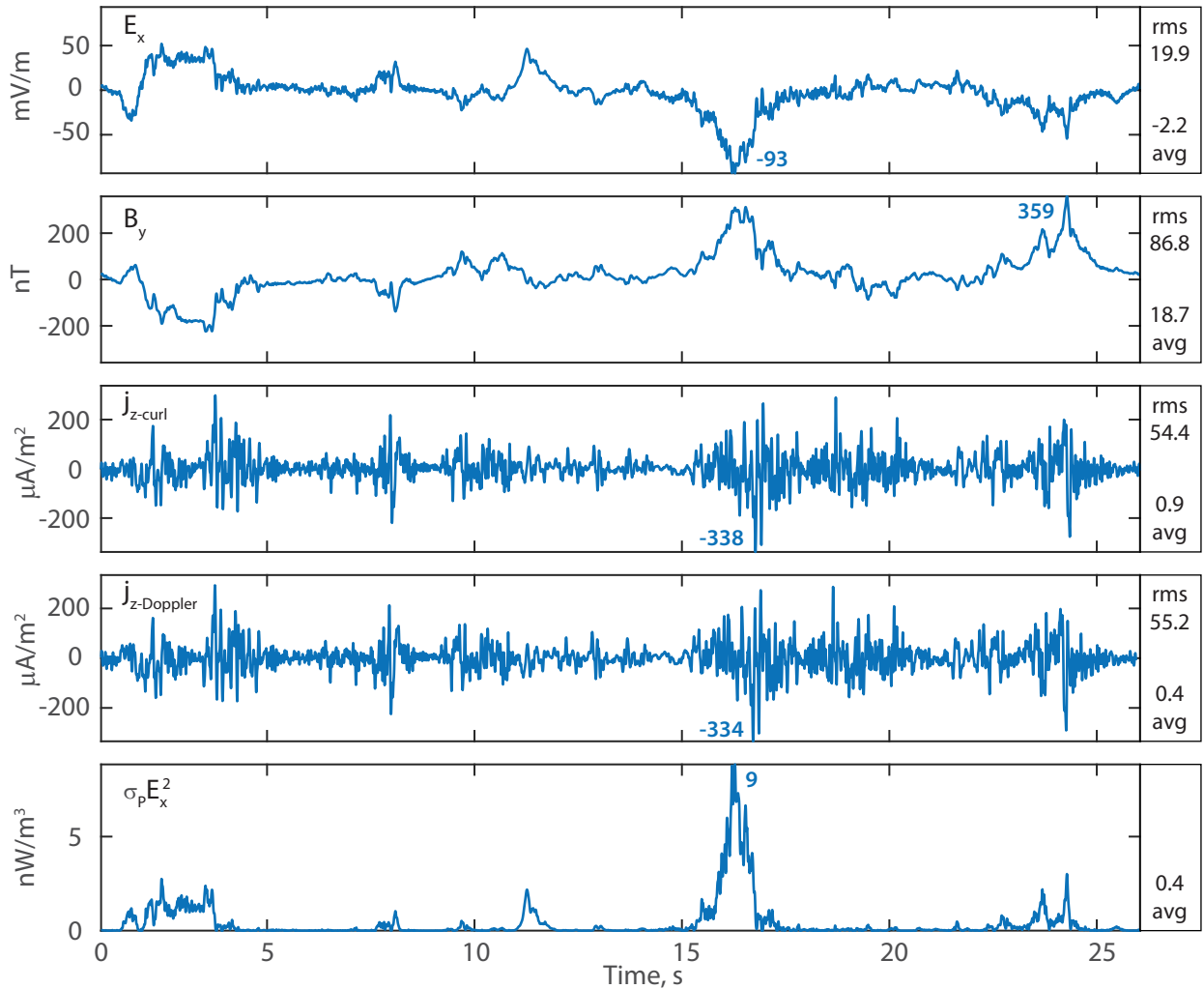

**Figure S3.** Illustrative synthetic time series of Alfvénic fields along a simulated CHAMP orbital segment through a 200-km wide (MLAT) cusp. The fields are calculated from (9) and (10) with $x = V_s t$. From top to bottom, MLAT component of the electric field $E_x$, MLT component of the magnetic field $B_y$, FAC calculated as the curl of $B_y$, FAC calculated as the Doppler derivative of $B_y$, and volumetric Joule heating rate $\sigma_P E_x^2$. Maximum value of each field is marked near the peak; rms and average values to the right.

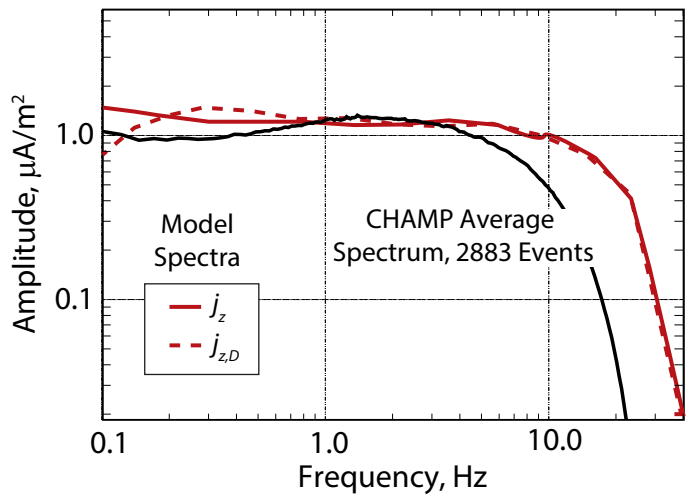

**Figure S4.** Octave-average spectra for the field-aligned current determined from the curl ($j_z$) and the Doppler-derivative ($j_{z,D}$) of $B_y(x, z, t)$ for the model ensemble of Alfvén waves and from CHAMP data averaged over 2883 events (Rother et al., 2007)[1]. The model spectra are calculated for average background parameters on 27 March 2003.

## Text S6. Altitudes profiles of V, $\sigma_P$ and $\sigma_0$ used to compute FDTD solutions

Figure S5 shows profiles for the wave speed $V$, Pedersen conductivity $\sigma_P$, and parallel conductivity $\sigma_0$ up to 2000 km altitude on 27 March 2003, calculated using IRI and MSIS to determine the ionospheric and thermospheric variables in equations (4)-(6). These parameters are documented in the repository as "iriday2parameters.txt" and "msisday2parameters.txt." The electron density profile obtained from IRI is subsequently multiplied by a constant so that the peak electron density, or NmF2, matches that of the average electron density profile in the central cusp from the baseline CMIT simulation for each day of the study. These profiles are essentially the same for 26 March 2003. Above 2000 km, $V$ and $\sigma_P$ are constant and equal to their respective values at 2000 km up to 5000 km. To facilitate convergence of the wave solver, $\sigma_0$ is interpolated to a large value specified at the 5000 km top of simulation space, here $10^{10}$ S/m. The results are not sensitive to this value as long as it is sufficiently large. The resulting profiles are used to calculate FDTD wave solutions as described in section 2.3 of the main text.

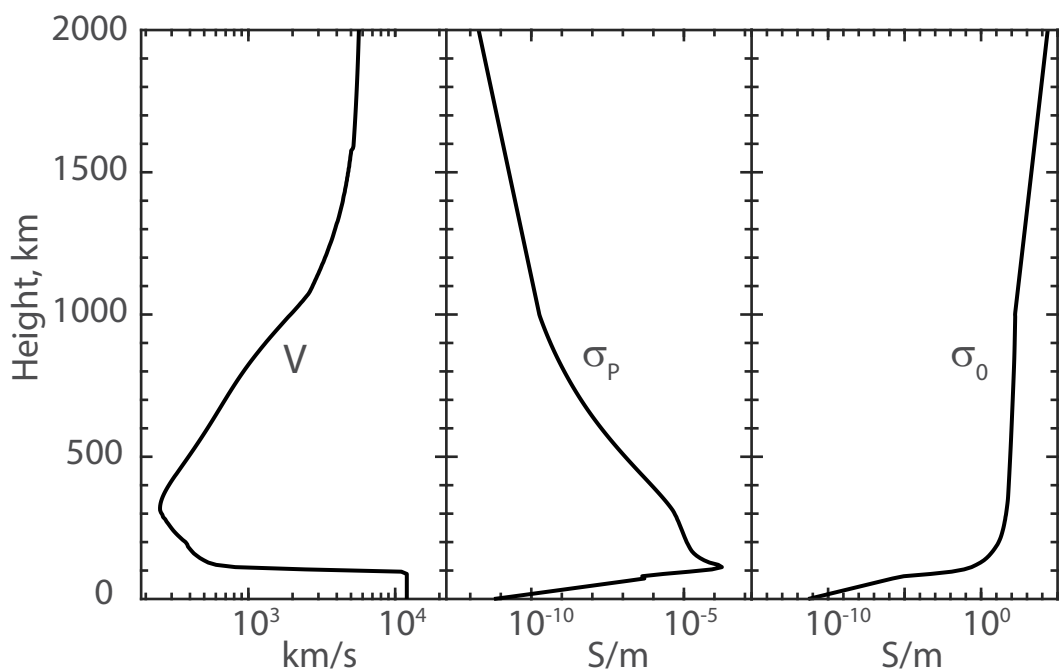

**Figure S5.** Altitude profiles up to 2000 km for the Alfvén speed ($V$), Pedersen ($\sigma_P$) and parallel ($\sigma_P$) conductivities on 27 March 2003 used by the FDTD solver.

## Text S7. Notes on electromagnetic energy deposition and frictional heating

The term Joule heating as conventionally used in the IT literature (and in most of the papers referenced in the Introduction to this paper) refers to the rate of IT electromagnetic energy dissipation calculated as

$$\mathbf{J} \cdot \mathbf{E}' = \sigma_P E_{\perp}'^2 + \sigma_0 E_{\parallel}^2 \ . \tag{S7.1}$$

$\mathbf{E}' = \mathbf{E} + \mathbf{u_n} \times \mathbf{B}$ is the electric field in the reference frame of the neutral gas moving with velocity $\mathbf{u_n}$, $\mathbf{B}$ is the background (geomagnetic) field, and $\sigma_P$ and $\sigma_0$ are the Pedersen and direct conductivities given in Sec. 2.2. The contribution $\sigma_0 E_{\parallel}^2$ to Joule heating is sometimes distinguished as Ohmic heating owing to the functional similarity between $\sigma_0$ and the electrical conductivity of a simple solid conductor. In most applications the dominant term in (S7.1) is $\sigma_P E_{\perp}^2$, which alone is often taken to be the Joule heating rate in IT studies.

Equation (S7.1) assumes the IT variability is quasistatic, i.e., it occurs on time scales much larger than both the gyroperiod and the mean time between collisions. This assumption allows an ionospheric Ohm's law to be expressed as

$$\mathbf{J} = \sigma_P \mathbf{E}'_\perp + \sigma_H \mathbf{b} \times \mathbf{E}' + \sigma_0 E_\parallel \mathbf{b} \, . \tag{S7.2}$$

From the perspective of kinetic theory, IT Joule and Ohmic heating arise from friction between the electron, ion and neutral gases. When the energy transfer integral for collisional interactions can be given in terms of Maxwell molecule collisions, the heating rate for species $r$ due to collisions with species $s$ with frequency $\nu_{rs}$ is[12]

$$\frac{\delta Q_{rs}}{\delta t} = \frac{m_r n_r \nu_{rs}}{m_r + m_s} \left[ 3 k_b \left( T_s - T_r \right) + m_s \left| \mathbf{v}_r - \mathbf{v}_s \right|^2 \right] . \tag{S7.3}$$

$T_s, \mathbf{v}_s, m_s$ and $n_s$ are the species temperature, velocity, atomic mass and number density; $k_B$ is Boltzmann's constant.

At altitudes below 400-500 km, the length scale $H$ for spatial variations and the time scale $\tau$ due to ion gyro motion $\Omega_i^{-1}$ and ion-neutral collisions $\nu_{in}^{-1}$ satisfy $|\mathbf{v}_i - \mathbf{u}_n| \tau \ll H$. In this case, collisional ion-neutral energy exchange achieves an approximate balance between temperature and friction coupling[13] with $3 k_b (T_i - T_n) \cong m_n |\mathbf{v}_i - \mathbf{u}_n|^2$. The heating rate of the neutral gas due to ion friction, $Q_J^{T_n}$ in the notation of Sec. 2.4, is then

$$Q_J^{T_r} = \sum_i \frac{\delta Q_{ni}}{\delta t} \cong \sum_i m_i n_i \nu_{in} \left| \mathbf{v}_i - \mathbf{u}_n \right|^2 . \tag{S7.4}$$

Substituting the ion-neutral velocity difference in the quasi-static limit,

$$\left| \mathbf{v}_i - \mathbf{u}_n \right|^2 = \frac{e^2}{m_i^2} \frac{E'^2}{\Omega_i^2 + \nu_{in}^2} \, , \tag{S7.5}$$

in (S7.4), the frictional heating rate of the neutral gas can then be equivalently calculated as the Joule heating rate in the neutral-wind frame:[14,15]

$$\begin{aligned} Q_J^{T_r} = \sum_i m_i n_i \nu_{in} \left| \mathbf{v}_i - \mathbf{u}_n \right|^2 &= \sum_i \frac{e^2 n_i}{m_i} \frac{\nu_{in}}{\Omega_i^2 + \nu_{in}^2} E'^2 \\ &= \sum_i \sigma_{Pi} \left| \mathbf{v}_{E \times B} - \mathbf{u}_n \right|^2 = \sigma_P E'^2 . \end{aligned} \tag{S7.6}$$

The partial Pedersen conductivities $\sigma_{Pi}$ are defined implicitly by the third equality in (S7.6). This particular form is used to calculate the Joule heating rate (né ion-neutral frictional heating) in the TIEGCM[16].

This result can also be derived in the MHD approximation in the center-of-mass frame for plasma and neutral constituents (essentially the neutral-gas frame),[14] when small corrections for Ohmic heating from the generalized Ohm's law and the work done on the neutrals in this frame are neglected.[17,18] Alternatively, the Joule heating rate in the ion reference frame is essentially the Ohmic heating rate in MHD [17,19] except at altitudes from about 120-300 km, where it differs by the additional friction term[18] $n_e m_e \nu_{en} |\mathbf{u}_{MHD} - \mathbf{u}_n|^2$. Ion-neutral friction appears as separate mechanical energy dissipation in the ion reference-frame formulation and is not associated with electromagnetic energy dissipation in that frame. These formal distinctions regarding neutral gas heating, whether considered as electromagnetic or mechanical energy dissipation, do not change the calculated rate of neutral gas heating or its physical origin arising from the collisional interaction between charged particles and the neutral gas.

In our treatment of Joule dissipation (Sec. 2.4), Ohmic dissipation is unimportant for the perpendicular wavelengths of interest and only conventional Joule dissipation is evaluated for Alfvénic energy deposition $Q_{J,A}$. We now show that (S7.6) can be used to evaluate collisional Alfvénic energy deposition with the result $Q_{J,A} = \sigma_P E_{A\perp}^2$, even though Alfvén wave dynamics do not fully satisfy the quasistatic approximation assumed in deriving (S7.6). This simple Joule heating formula is applicable to Alfvén dynamics because, as discussed in SI Text S3, at sufficiently high wave frequency and low ion-neutral collision frequency, the ratio of the perturbed neutral gas velocity to Alfvén wave ion velocity is negligibly small.

Since Lysak's formulation[8] is the basis for equations (1)-(3) in Sec. 2.2, we start with his equation (A3) for the linear velocity perturbation for ionized species $s$:

$$\mathbf{v}_{s\perp} = \frac{e}{m_s \varpi_s^2} \left[ \left( \frac{\partial}{\partial t} + \nu_s \right) \mathbf{E}_\perp + \mathbf{E}_\perp \times \mathbf{\Omega}_s \right]. \tag{S7.7}$$

We defined $\varpi_s^2 \equiv \nu_s^2 + \Omega_s^2$. As discussed by Lysak, this expression with $\partial/\partial t \to \omega$ assumes $\omega^2, \omega\nu_s \ll \Omega_s^2$. Retaining corrections with $\omega\nu_s \geq \Omega_s^2$ improves the accuracy of the polarization drift in the $E$ layer (term proportional to $\partial/\partial t$), but the resulting differences in Alfvén wave propagation are minor because the polarization current in the $E$ layer is much less than the Pedersen current.[20] Equation (S7.7) also neglects the small, wave-induced perturbation in neutral velocity stimulated by ion-neutral friction (see SI Text S3). From (S7.7), we obtain

$$v_{s\perp}^2 = \frac{e^2}{m_s^2 \varpi_s^2} \left| \left( \frac{\partial}{\partial t} + \nu_s \right) \mathbf{E}_\perp + \mathbf{E}_\perp \times \mathbf{\Omega}_i \right|^2 = \frac{e^2}{m_s^2 \varpi_s^4} \left[ \left| \frac{\partial \mathbf{E}_\perp}{\partial t} \right|^2 + 2\nu_s \mathbf{E}_\perp \cdot \frac{\partial \mathbf{E}_\perp}{\partial t} + \varpi_s^2 E_\perp^2 \right]$$

$$\simeq \frac{\Omega_s^2}{\varpi_s^2} \frac{E_\perp^2}{B_0^2} \tag{S7.8}$$

where $\omega^2, \omega\nu_s \ll \Omega_s^2$ has been used again in the last equality. In light of the analysis in SI Text S3 we can neglect $\mathbf{u_n}$ compared with $\mathbf{v}_i$ in $\mathbf{E'}_\perp$ in (S7.6) to obtain for the Joule heating rate due to Alfvén-wave energy deposition

$$Q_{J,A} \simeq \sum_i m_i n_i \nu_i v_{iA\perp}^2 = \sum_i m_i n_i \nu_i \frac{\Omega_i^2}{\varpi_i^2} \frac{E_{A\perp}^2}{B_0^2} = \sum_i \frac{n_i e^2}{m_i} \frac{\nu_i}{\varpi_i^2} E_{A\perp}^2 = \sigma_P E_{A\perp}^2. \tag{S7.9}$$